\documentclass[%
reprint,
amsmath,amssymb,
aps,
]{revtex4-2} 
\usepackage{graphicx}% Include figure files
\usepackage{dcolumn}% Align table columns on decimal point
\usepackage{bm}% bold math
\usepackage[mathlines]{lineno}% Enable numbering of text and display math
\usepackage{amsmath}
\usepackage{multirow}
\begin{document}
	
	\title[title]{X-ray absorption of cold gas: Simulating interstellar molecular clouds in the laboratory}
	
	\author{I. Gissis}
	\email{itaigiss@gmail.com}
	\author{A. Fisher}
	\email{amnon.fisher@gmail.com}
	\author{E. Behar}
	\email{behar@physics.technion.ac.il}
	
	\affiliation{Physics Department, Technion, Israel Institute of Technology, Haifa 3200003, Israel}
	
	\date{\today}% It is always \today, today,
	
	\begin{abstract}
		Galactic and extra-galactic sources produce X-rays that are often absorbed by molecules and atoms in giant molecular clouds (GMCs), which provides valuable information about their composition and physical state. We mimic this phenomenon with a laboratory Z-pinch X-ray source, which is impinged on neutral molecular gas. The novel technique produces a soft X-ray pseudo continuum using a pulsed-current generator. The absorbing gas is injected from a 1\,cm long planar gas-puff without any window or vessel along the line of sight. An X-ray spectrometer with a resolving power of $\lambda/\Delta\lambda\sim$420, comparable to that of astrophysical space instruments, records the absorbed spectra. This resolution clearly resolves the molecular lines from the atomic lines; therefore, motivating the search of molecular signature in astrophysical X-ray spectra. The experimental setup enables different gas compositions and column densities. K-shell spectra of CO$_2$, N$_2$ and O$_2$ reveal a plethora of absorption lines and photo-electric edges measured at molecular column densities between $\sim$10$^{16}$\,cm$^{-2}$ -- 10$^{18}$\,cm$^{-2}$ typical of GMCs. We find that the population of excited-states, contributing to the edge, increases with gas density.		
	\end{abstract}
	
	%Use showkeys class option if keyword
	\maketitle
		
	\section{\label{sec:Introduction}Introduction}
	
	X-rays from X-ray binaries (XRBs) and active galactic nuclei (AGNs) are often absorbed in the interstellar (ISM) and intergalactic (IGM) media. Giant molecular clouds (GMC) are abundant in the ISM and in star-forming regions. One of the main questions relevant to GMCs is their chemistry and heating mechanism affected by X-rays, shock-waves and cosmic rays \cite{Meijerink_2005}. In X-ray dominated regions (XDR), the X-rays penetrate deep into the cloud, releasing secondary electrons, which heat, dissociate and further ionize the gas. IR and mm emission of molecular tracers provide temperature and composition diagnostics \cite{Tielens_2013,Maloney_1996} including for high red-shift AGNs \cite{Vallini_2018}. Complementary diagnostics comes from high-resolution X-ray (0.3 -- 10\,keV) absorption \cite{Joachimi_2016} and fluorescence \cite{Kawamuro_2020}. Thus, it is important to study the X-ray absorption of these tracers in the laboratory. 
	
	\subsection{\label{sec:LabMolecularAbsorption}Laboratory molecular absorption}
	Near edge X-ray absorption fine structure (NEXAFS) is a common method to characterize the elemental composition of thin solid samples, by probing the fine absorption features near the photo-electric edge. NEXAFS is originally and routinely performed in large beam facilities, i.e. synchrotrons \cite{Kincaid_1975} or free electron lasers \cite{Hiroshi_2020}. It is of high interest to various scientific disciplines such as the study of organic sheets \cite{Hahner_2006}, nanoscale samples \cite{HemrajBenny_2006}, inorganic polymers \cite{Dhez_2003}, and astrophysical dust \cite{Costantini_2019,Lee_2009}. Table-top NEXAFS systems that can carry out a large volume of tests at a high repetition rate are therefore desirable. Laser produced plasma (LPP) is the leading technique \cite{Wachulak_2018}. However, table-top LPPs require hundreds of shots to obtain a single high SNR spectrum. 
	We have developed the GLIDER \cite{Gissis_Glider2020}, a compact pulsed-power X-ray source of $\sim$10$^{17}$\,erg\,s$^{-1}$, which enables the measurement of NEXAFS of any matter-phase in an individual $\sim$10\,ns shot and dozens of shots per hour. In this paper, we report on NEXAFS diagnostics of astrophysical molecules found in GMCs.
	
	\section{\label{sec:ExperimentalSetup}Experimental setup and methodology}
\begin{figure}
		\includegraphics{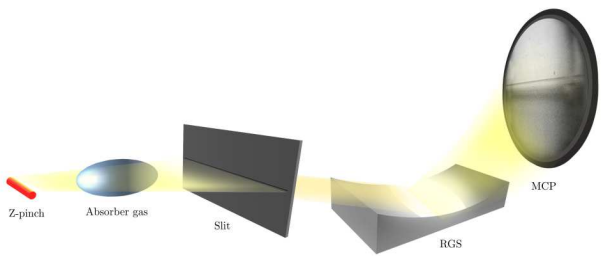}% Here is how to import EPS art
		\caption{\label{fig:OpticalSystem}Schematic experimental setup.}
	\end{figure}

	The experimental setup is depicted in Figure \ref{fig:OpticalSystem}. A gas-puff Z-pinch plasma X-ray source is produced using the GLIDER generator \cite{Gissis_Glider2020}. The Z-pinch source is optimized for absorption measurements of the CNO K-shell spectral range (20 -- 50\,\AA{}) by using imploding Kr gas. Unlike low-Z elements, Kr has a multitude of blended M-shell lines which produce a pseudo continuum soft X-ray and UV back-lighter to be absorbed by the CNO gas.
	The absorbing gas is injected through a second gas-puff along the line of sight, $\sim$300\,mm away from the pinch, and decoupled from it. The gas-puff injector is fast enough to prevent interaction between the absorber and the source. A reflection grating spectrometer (RGS) combined with a gated micro channel plate (MCP) detector records the absorption spectrum. Importantly, we avoid polymers such as polyimid or mylar windows, for they comprise C, N and O, which absorb strongly in the spectral range of interest. Indeed, our high resolution spectra of polyimid revealed significant C NEXAFS features. 
	To avoid ambiguous diagnostics and to still protect the grating from debris and stray light, a glass capillary array is placed between the gas absorber and the grating entrance slit. 
	
	\subsection{\label{sec:WavelengthCalibration}Wavelength calibration}
	Wavelength calibration was achieved using accurately known emission lines of highly ionized K-shell C, N and O in dedicated Z-pinch shots with no absorber. To calibrate the O K-edge around 23\,\AA{}, the resonance lines of N\textsuperscript{+5,+6} and O\textsuperscript{+6,+7} are used. To calibrate the N K-edge around 31\,\AA{} the resonant lines of C\textsuperscript{+4,+5} and N\textsuperscript{+5,+6} are used. The C K-edge around 42\,\AA{} region was calibrated using both $1^{st}$ order diffraction C lines, e.g. at 33.736 (Ly$\alpha$), 40.2678 (He$\alpha$), 34.973 (He$\beta$) and $2^{nd}$ order O lines, e.g. at 37.9340 (Ly$\alpha$), 43.202 (He$\alpha$).  
	Each $\sim$10\,\AA{} range is calibrated using 5--10 peaks resulting in systematic wavelength uncertainty of $\leq7$\,m\AA{}.
	The spectrometer alignment is sensitive to temperature fluctuations. Therefore, spectral calibration was performed within minutes before or after the absorption shots. 
		
	\subsection{\label{sec:TransSpectra}Transmission spectral modeling}
	Transmission spectra are obtained by dividing an absorbed spectrum by one in a separate shot, which is un-absorbed in the range of interest. 
	The transmission is defined as $e^{-\tau}$, where $\tau$ is the optical depth. The column density $N$ of the absorber is related to the optical depth through: $\tau=\,N\sigma$ where, $\sigma(\lambda)$ is the cross-section profile for absorption in a spectral line or an edge. Column densities are estimated by dividing the measured $\tau(\lambda)$ by the published $\sigma(\lambda)$ \cite{Berkowitz_2015}. The column density is also: $N=\int{n(l)\,dl}$, where $n(l)$ is the number density along the line of sight and is measured independently as described in Section \ref{sec:PressureGauge}.
	
	Each NEXAFS is modeled with a single edge and several resolved lines. The edge is modeled with a $\lambda^3$ profile (hydrogenic approximation) and the lines are modeled as Gaussians with fitted widths. Since we did not observe ionized or highly excited states, we assume all molecules are in their electronic ground-state, but could be vibrationally or rotationally excited. These excitations may be reflected in the absorption lines, but the edge still manifests the total molecular column density. 
	The entire model was further convoluted with a Gaussian of $\sigma=50$ -- 70\,m\AA{} accounting for the instrumental broadening. The data are fitted using NASA's Xspec spectral analysis code \cite{Arnaud_1996}.
	
	\section{\label{sec:Results}Results}
\begin{figure*}
		\includegraphics[width=\textwidth]{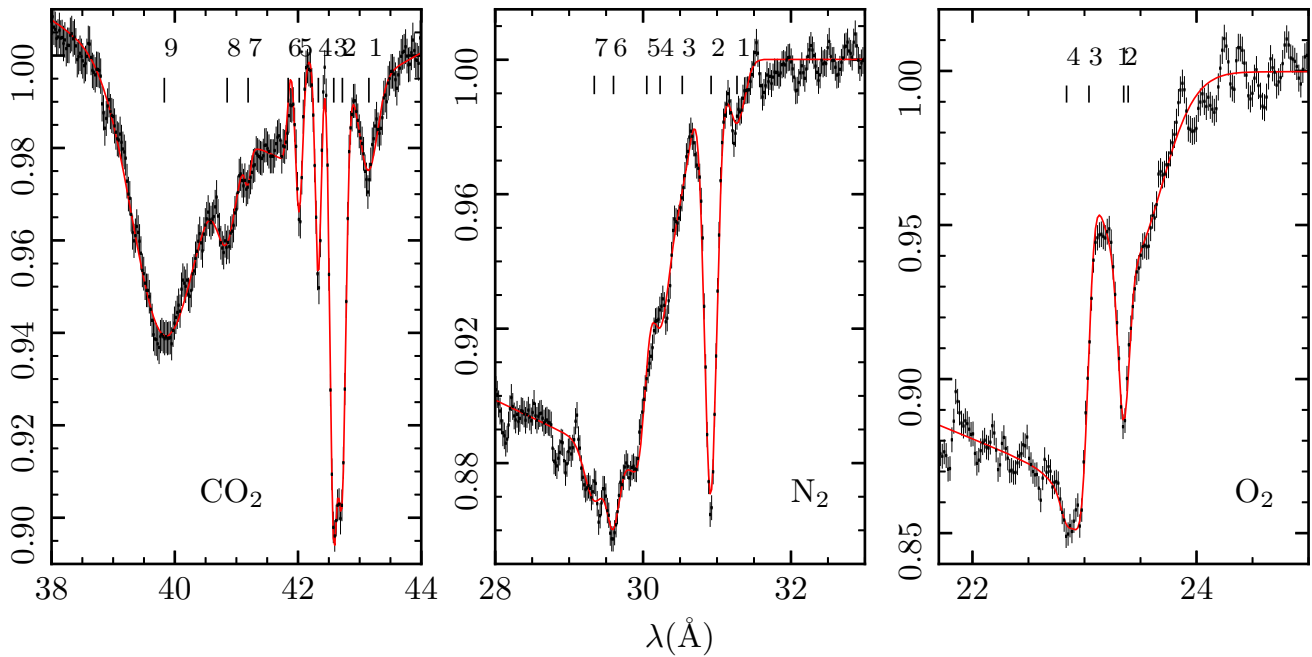}
		\caption{\label{fig:specThree}Transmission spectra of CO$_2$ (C edge), N$_2$ and O$_2$ over a Kr back-lighter. The best fit model is plotted in solid red. The resolved transitions are marked on the figure and detailed in Tables \ref{tab:CO2Lines},\ref{tab:N2Lines} and \ref{tab:O2Lines}.}
	\end{figure*}

	In Figure \ref{fig:specThree} we present normalized NEXAFS spectra of CO$_2$, N$_2$ and O$_2$. The models capture all main features identified in the literature to within $\sim$1\,eV. The results are summarized in Tables \ref{tab:CO2Lines}, \ref{tab:N2Lines}, and \ref{tab:O2Lines}. The molecular neutral K$\alpha$ and the edge is clearly resolved in all three spectra. No ionized spectral lines are found short-ward of the edge. In C and N the neutral atom K$\alpha$ is resolved and is $\sim$10 times weaker than the molecular K$\alpha$. Hence, the molecules are essentially neutral and only mildly dissociated.  
	 
	A few exceptional details are worth noting: 
	The molecular K$\alpha$ in all CO$_2$ shots is split into two peaks (No. 2 and 3 in Figure \ref{fig:specThree}). The mean peak value is 42.63\,\AA{}, consistent with the $(2\sigma_g)^{-1}2\pi_u$ transition \cite[Figure 4.2.5 therein]{Berkowitz_2015}. The deep feature at 39.8\,\AA{} (No. 9) dominates the edge as it does in \cite{Sham_1989}. The edge starts at 41.84\,\AA{} and its recovery is seen shortward of 39\,\AA{}.
	
	\begin{table}[b]
		\caption{\label{tab:CO2Lines}
			CO$_2$ absorption lines and photo-electric edge around the C K-edge and their $1\sigma$ confidence interval.}
		\begin{ruledtabular}
			\begin{tabular}{lllll}
				No.&Central wavelength&Central energy&Literature&Assignment\\
				&\AA & eV & eV & \\
				\hline
				1& $43.15_{-0.01}^{+0.02}$ & $287.4_{-0.3}^{+0.1}$& 285 \footnotemark[1]& $1s2s^22p$  \\
				2& $42.72_{-0.01}^{+0.02}$ & $290.3_{-0.3}^{+0.1}$ & \multirow{2}{*}{290.77 \footnotemark[2]}& \multirow{2}{*}{$(2\sigma_g)^{-1}2\pi_u$}\\
				3& $42.58_{-0.01}^{+0.02}$ & $291.3_{-0.1}^{+0.1}$&  &   \\
				4& $42.33_{-0.01}^{+0.02}$ & $293.0_{-0.1}^{+0.1}$& 292.74 \footnotemark[2]&  $(2\sigma_g)^{-1}5\sigma_g$ \\
				5& $42.02_{-0.01}^{+0.01}$ & $295.1_{-0.1}^{+0.1}$& 295 \footnotemark[3]& $3p$  \\
				6& $41.84_{-0.02}^{+0.01}$ & $296.4_{-0.1}^{+0.1}$& 297.6 \footnotemark[3]& IP \\
				7& $41.19_{-0.02}^{+0.01}$ & $301.1_{-0.1}^{+0.1}$& 300.5--301 \footnotemark[3]& shake-up \\
				8& $40.85_{-0.04}^{+0.01}$ & $303.6_{-0.1}^{+0.2}$& 304.2 \footnotemark[3]& shake-up  \\
				9& $39.83_{-0.04}^{+0.01}$ & $311.4_{-0.1}^{+0.2}$& 311--314\footnotemark[2]$^,$\footnotemark[3]&shake-up\footnotemark[4]\\		 			 
			\end{tabular}
		\end{ruledtabular}
		\footnotetext[1]{Ref.\cite{Hasoglu_2010} - mean value for all transitions}		
		\footnotetext[2]{Ref.\cite{Berkowitz_2015}}
		\footnotetext[3]{Ref.\cite{Sham_1989}}
		\footnotetext[4]{Blended with $\sigma^*$ shape-resonance}
	\end{table}	

	In N$_2$ as well, almost all the absorption features fit their tabulated positions to within $\sim$1\,eV. An exception is the edge, which is measured here $\sim$3\,eV above the value in \cite{Berkowitz_2015}. This is likely due to blending of the edge with the strong double-excitation feature at $\sim$413\,eV \cite[Figure 3.2.6 therein]{Berkowitz_2015}. 
	
	\begin{table}[b]
	\caption{\label{tab:N2Lines}
		Measured N$_2$ absorption lines and photo-electric edge and their $1\sigma$ confidence interval.}
	\begin{ruledtabular}
		\begin{tabular}{lcccl}
			No.&$\lambda$&E&Literature&Assignment\\
			&\AA & eV & eV & \\
			\hline
			 1& $31.27_{-0.01}^{+0.02}$ & $396.6_{-0.3}^{+0.2}$& 396 \footnotemark[1]& 1s  \\
			 2& $30.92_{-0.01}^{+0.02}$ & $401.1_{-0.2}^{+0.2}$& 400.96 \footnotemark[2]& $2p\pi_g(\pi^*)$ \\
			 3& $30.53_{-0.02}^{+0.02}$ & $406.2_{-0.2}^{+0.2}$& 406.13 \footnotemark[2]& $3s\sigma_g$\\
			 4& $30.23_{-0.02}^{+0.01}$ & $410.3_{-0.2}^{+0.2}$& 408.36 \footnotemark[2]& $4s\sigma_g/3d$  \\
			 
			 \multirow{2}{*}{5}& \multirow{2}{*}{$30.05_{-0.02}^{+0.01}$}& \multirow{2}{*}{$412.8_{-0.2}^{+0.2}$}& 413\footnotemark[3] & Double excitation \\	
			 &  & & 409.9 \footnotemark[3]& IP\\		 
			 6& $29.60_{-0.02}^{+0.01}$ & $419.0_{-0.2}^{+0.2}$& 418.9 \footnotemark[2]& $\sigma_u,l=3$\\
			 7& $29.34_{-0.02}^{+0.01}$ & $422.7_{-0.2}^{+0.3}$& 421 \footnotemark[2]& shake-up 		 			 
		\end{tabular}
	\end{ruledtabular}
	\footnotetext[1]{Ref.\cite{Garcia_2009}}
	\footnotetext[2]{Ref.\cite{Hitchcock_1980}}
	\footnotetext[3]{Ref.\cite{Berkowitz_2015}} 	
	\end{table}

    In O$_2$ the K$\alpha$ resonance comprises a few transitions. We fit it with two blended peaks at 23.37\,\AA{}, A broad one (No. 1 in Figure \ref{fig:specThree}) and a narrow one (No. 2). Unlike N$_2$ and CO$_2$, in O$_2$ the atomic O K$\alpha$ at $\sim$527\,eV \cite{McLaughlin_2013} is absent and likely obscured by the broad No. 1 feature. Our confidence in the O$_2$ interpretation is lower than in N$_2$ or CO$_2$ for the Kr emission in this energy range is not as smooth, and residual Kr emission may linger in the transmission spectrum.
    
    \begin{table}[b]
	\caption{\label{tab:O2Lines}
		O$_2$ absorption lines and photo-electric edge and their $1\sigma$ confidence interval.}
	\begin{ruledtabular}
		\begin{tabular}{lcccl}
			No.&$\lambda$&$E$&Literature&Assignment\\
			&\AA & eV & eV & \\
			\hline
			1& $23.39_{-0.01}^{+0.02}$ & $530.3_{-0.5}^{+0.2}$&\multirow{2}{*}{530.8 \footnotemark[1]}&\multirow{2}{*}{$\pi^*$}  \\
			2& $23.35_{-0.01}^{+0.02}$ & $531.0_{-0.2}^{+0.2}$& &  \\
			3& $23.04_{-0.01}^{+0.01}$ & $538.3_{-0.2}^{+0.3}$& $\sim$540.5\footnotemark[2]& IP\\
			4& $22.84_{-0.03}^{+0.01}$ & $543.0_{-0.2}^{+0.4}$& 539.3\footnotemark[2]&$\sigma^*$ \\			 			 
		\end{tabular}
	\end{ruledtabular}
	\footnotetext[1]{Ref.\cite{Hitchcock_1980}. Lines 1 and 2 are blended, while line 1 is broadened (Figure \ref{fig:specThree}).}
	\footnotetext[2]{Ref.\cite{Berkowitz_2015}. We measure a broad $\sigma^*$ shape-resonance, next to the edge.}	
	\end{table}	
		 
	\subsection{\label{sec:ColumnDensity}Column density measurements}
	%The column density can be deduced from the fitted optical depth as a function of $\lambda$ according to $N=\tau/\sigma$, where $\sigma$ was taken from \cite{Berkowitz_2015}. 
	For the sample spectra of Figure \ref{fig:specThree} we obtained $1.2\pm0.1, 10\pm2, 11\pm2$ in units of $10^{16}$\,cm$^{-2}$ for the CNO edges. This 10\%--20\% uncertainty is attributed to the fitting process of the $\sigma(\lambda)$ profile to the measured $\tau(\lambda)$. 

\begin{figure}
		\includegraphics{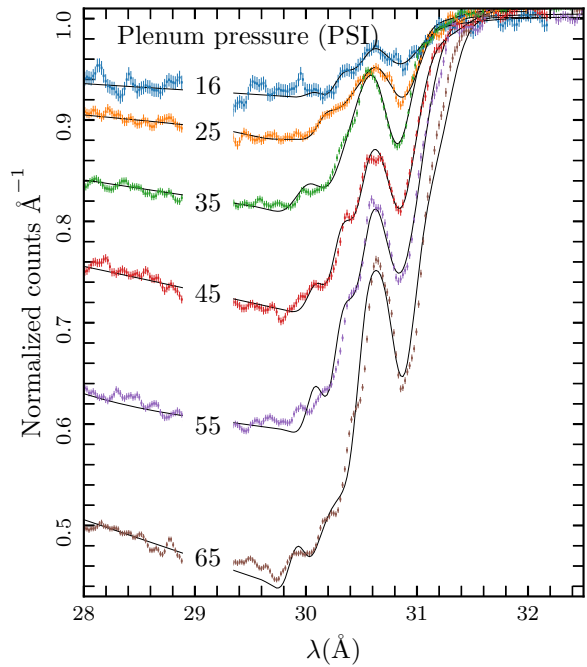}
		\caption{\label{fig:N2Pressures}Spectra of N$_2$ absorber on a Kr pinch background. The absorber puff plenum pressure in each shot is written on its corresponding spectrum.}
	\end{figure}

	A significant advantage of our experiment is the flexibility to control the absorber density and therefore the column density. In Figure \ref{fig:N2Pressures} we present a set of experiments with N$_2$ at different initial plenum pressures. It can be seen from the figure that all NEXAFS features deepen as the pressure (and density) is increased.
	
\begin{figure}[t!]
		\includegraphics{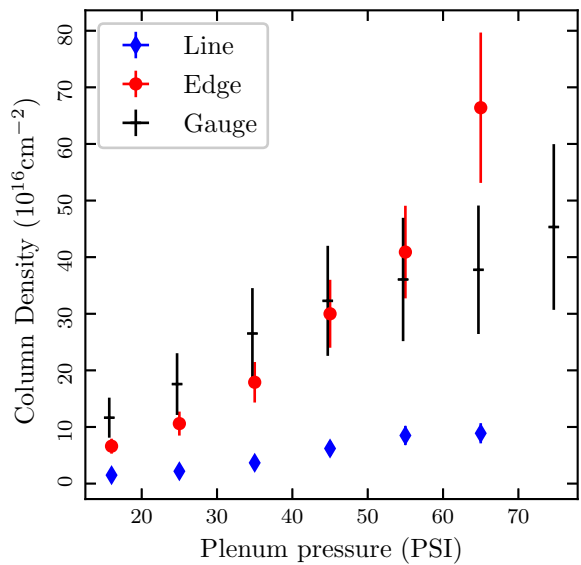}
		\caption{\label{fig:colDens}Column density as a function of the absorber puff plenum pressure. Measurements are based on the line, edge and a piezo-electric pressure gauge. The column density uncertainties reflect also the variance between shots.}
	\end{figure}

	The measured N$_2$ column densities are presented in Figure \ref{fig:colDens} as a function of the pressure, separately for the K$\alpha$ resonance (blue diamonds) and the edge (red circles). The values deduced from the edge are consistently higher than those deduced by the line. This ratio gradually increases with the plenum pressure. The N$_2$ edge cross-section is enhanced by additional features of excited molecules \cite[Figure 3.2.6 therein]{Berkowitz_2015}. This increases the opacity at the edge and thus, the deduced column density by more than a factor of 2. The remaining discrepancy between the edge and the line column density could be explained by the lines being absorbed only by ground-state molecules, while the edge is produced also by excited molecules in the ground configuration, and this excitation increases with pressure. Supporting evidence is the strong shake-up features measured in both  CO$_2$ and N$_2$ (See Tables \ref{tab:CO2Lines}, \ref{tab:N2Lines}) following the X-ray and UV excitation, which requires a broad-band source similar to astrophysical sources. 
	
	The lack of significant heating and dissociation resembles less an XDR and more a remote GMC. This may be due to the relatively short X-ray pulse of $\sim$10\,ns, compared to the long dissociation times of $\sim$$\mu$s. The preceding UV pulse is also short, $\sim$30\,ns. The dissociation time $t_{\mathrm{diss}}=\left(F\sigma\right)^{-1}$, where $F=10^{23}$\,ph\,s$^{-1}$\,cm$^{-2}$ is the measured photon flux above 10\,eV \cite{Gissis_Glider2020}. The cross-section is $\sigma=10^{-18} - 10^{-16}$\,cm$^2$ varying over the spectral band\cite{Huebner_2015}, which results in $t_{\mathrm{diss}}=0.1 - 10\,\mu$s.

	\subsection{\label{sec:PressureGauge}Independent density measurement}

	The pressure of the absorber was measured independently with a piezo-electric gauge, located downstream from the absorber puff nozzle exit, at the vertical position of the spectrometer slit. The gauge measures the stagnation pressure which has a fixed ratio to the isentropic pressure and to the number density. The column density is this number density times the known planar nozzle length of 1\,cm.
	The deduced gas column density is plotted in Figure \ref{fig:colDens} as the black crosses. The agreement between the spectroscopically deduced column densities and the gauge measurement enhances our confidence in the cross-sections used, albeit the experimental uncertainties.   
	
	Beyond the 10\%--20\% uncertainty in the column density measured in an individual shot, the transmission spectra (Figures \ref{fig:specThree} and \ref{fig:N2Pressures}) vary from shot to shot by $\sim$15\%. These variations between shots stem from the imperfect mechanics of the absorber puff, and are reflected in the error bars in Figure \ref{fig:colDens}.	The gauge density uncertainty results from the statistical repeatability of the injected gas density and the gauge SNR. All sum up to 30\% uncertainty.  
	
	\section{\label{sec:Conclusions}Conclusions}
	We present a novel experiment to measure X-ray absorption by gaseous abundant molecules and atoms, mimicking astrophysical absorption in the ISM. This is the first time absorption of abundant astrophysical gases is measured with a pulsed-power X-ray source. We measured high resolution NEXAFS spectra of CO$_2$, N$_2$ and O$_2$, identifying numerous absorption lines. In the present experiment the gas is excited by a broad-band pseudo-continuum Kr Z-pinch source. Therefore, the absorber is photo-excited similarly to molecular clouds exposed to X-ray astrophysical sources. We demonstrate a wide dynamic range of absorber densities. Utilizing the X-ray astronomy fitting code Xspec, we deduced column densities of $\sim$10$^{17}$\,cm$^{-2}$, in good agreement with independently measured pressure. 
	 
	The present setup opens the path for many more high-resolution X-ray absorption experiments that can support spectral interpretation of astrophysical sources and material science research. Improving the accuracy of the independent density measurement will allow to benchmark excitation cross-sections. We are now preparing to integrate a second spectrometer and measure the absorbed and un-absorbed spectra simultaneously; thus, improving the accuracy of the NEXAFS spectrum. We are currently extending our measurements to photo-ionized atomic species, also abundant in the ISM.
	
	\begin{acknowledgments}
	We acknowledge support by the Center of Excellence of the ISF (grant No. 2752/19), and a grant from the Pazy Foundation. We thank S. Bialy for useful comments on the manuscript.
	\end{acknowledgments}
	
	\bibliography{XrayAbsorption_Ver1}% Produces the bibliography via BibTeX.

\begin{thebibliography}{23}%
\makeatletter
\providecommand \@ifxundefined [1]{%
 \@ifx{#1\undefined}
}%
\providecommand \@ifnum [1]{%
 \ifnum #1\expandafter \@firstoftwo
 \else \expandafter \@secondoftwo
 \fi
}%
\providecommand \@ifx [1]{%
 \ifx #1\expandafter \@firstoftwo
 \else \expandafter \@secondoftwo
 \fi
}%
\providecommand \natexlab [1]{#1}%
\providecommand \enquote  [1]{``#1''}%
\providecommand \bibnamefont  [1]{#1}%
\providecommand \bibfnamefont [1]{#1}%
\providecommand \citenamefont [1]{#1}%
\providecommand \href@noop [0]{\@secondoftwo}%
\providecommand \href [0]{\begingroup \@sanitize@url \@href}%
\providecommand \@href[1]{\@@startlink{#1}\@@href}%
\providecommand \@@href[1]{\endgroup#1\@@endlink}%
\providecommand \@sanitize@url [0]{\catcode `\\12\catcode `\$12\catcode
  `\&12\catcode `\#12\catcode `\^12\catcode `\_12\catcode `\%12\relax}%
\providecommand \@@startlink[1]{}%
\providecommand \@@endlink[0]{}%
\providecommand \url  [0]{\begingroup\@sanitize@url \@url }%
\providecommand \@url [1]{\endgroup\@href {#1}{\urlprefix }}%
\providecommand \urlprefix  [0]{URL }%
\providecommand \Eprint [0]{\href }%
\providecommand \doibase [0]{https://doi.org/}%
\providecommand \selectlanguage [0]{\@gobble}%
\providecommand \bibinfo  [0]{\@secondoftwo}%
\providecommand \bibfield  [0]{\@secondoftwo}%
\providecommand \translation [1]{[#1]}%
\providecommand \BibitemOpen [0]{}%
\providecommand \bibitemStop [0]{}%
\providecommand \bibitemNoStop [0]{.\EOS\space}%
\providecommand \EOS [0]{\spacefactor3000\relax}%
\providecommand \BibitemShut  [1]{\csname bibitem#1\endcsname}%
\let\auto@bib@innerbib\@empty
%</preamble>
\bibitem [{\citenamefont {{Meijerink}}\ and\ \citenamefont
  {{Spaans}}(2005)}]{Meijerink_2005}%
  \BibitemOpen
  \bibfield  {author} {\bibinfo {author} {\bibfnamefont {R.}~\bibnamefont
  {{Meijerink}}}\ and\ \bibinfo {author} {\bibfnamefont {M.}~\bibnamefont
  {{Spaans}}},\ }\bibfield  {title} {\bibinfo {title} {{Diagnostics of
  irradiated gas in galaxy nuclei. I. A far-ultraviolet and X-ray dominated
  region code}},\ }\href {https://doi.org/10.1051/0004-6361:20042398}
  {\bibfield  {journal} {\bibinfo  {journal} {Astronomy and Astrophysics}\
  }\textbf {\bibinfo {volume} {436}},\ \bibinfo {pages} {397} (\bibinfo {year}
  {2005})},\ \Eprint {https://arxiv.org/abs/astro-ph/0502454}
  {arXiv:astro-ph/0502454 [astro-ph]} \BibitemShut {NoStop}%
\bibitem [{\citenamefont {{Tielens}}(2013)}]{Tielens_2013}%
  \BibitemOpen
  \bibfield  {author} {\bibinfo {author} {\bibfnamefont {A.~G.~G.~M.}\
  \bibnamefont {{Tielens}}},\ }\bibfield  {title} {\bibinfo {title} {{The
  molecular universe}},\ }\href {https://doi.org/10.1103/RevModPhys.85.1021}
  {\bibfield  {journal} {\bibinfo  {journal} {Reviews of Modern Physics}\
  }\textbf {\bibinfo {volume} {85}},\ \bibinfo {pages} {1021} (\bibinfo {year}
  {2013})}\BibitemShut {NoStop}%
\bibitem [{\citenamefont {{Maloney}}\ \emph {et~al.}(1996)\citenamefont
  {{Maloney}}, \citenamefont {{Hollenbach}},\ and\ \citenamefont
  {{Tielens}}}]{Maloney_1996}%
  \BibitemOpen
  \bibfield  {author} {\bibinfo {author} {\bibfnamefont {P.~R.}\ \bibnamefont
  {{Maloney}}}, \bibinfo {author} {\bibfnamefont {D.~J.}\ \bibnamefont
  {{Hollenbach}}},\ and\ \bibinfo {author} {\bibfnamefont {A.~G.~G.~M.}\
  \bibnamefont {{Tielens}}},\ }\bibfield  {title} {\bibinfo {title}
  {{X-Ray--irradiated Molecular Gas. I. Physical Processes and General
  Results}},\ }\href {https://doi.org/10.1086/177532} {\bibfield  {journal}
  {\bibinfo  {journal} {The Astrophysical Journal}\ }\textbf {\bibinfo {volume}
  {466}},\ \bibinfo {pages} {561} (\bibinfo {year} {1996})}\BibitemShut
  {NoStop}%
\bibitem [{\citenamefont {{Vallini}}\ \emph {et~al.}(2018)\citenamefont
  {{Vallini}}, \citenamefont {{Pallottini}}, \citenamefont {{Ferrara}},
  \citenamefont {{Gallerani}}, \citenamefont {{Sobacchi}},\ and\ \citenamefont
  {{Behrens}}}]{Vallini_2018}%
  \BibitemOpen
  \bibfield  {author} {\bibinfo {author} {\bibfnamefont {L.}~\bibnamefont
  {{Vallini}}}, \bibinfo {author} {\bibfnamefont {A.}~\bibnamefont
  {{Pallottini}}}, \bibinfo {author} {\bibfnamefont {A.}~\bibnamefont
  {{Ferrara}}}, \bibinfo {author} {\bibfnamefont {S.}~\bibnamefont
  {{Gallerani}}}, \bibinfo {author} {\bibfnamefont {E.}~\bibnamefont
  {{Sobacchi}}},\ and\ \bibinfo {author} {\bibfnamefont {C.}~\bibnamefont
  {{Behrens}}},\ }\bibfield  {title} {\bibinfo {title} {{CO line emission from
  galaxies in the Epoch of Reionization}},\ }\href
  {https://doi.org/10.1093/mnras/stx2376} {\bibfield  {journal} {\bibinfo
  {journal} {Monthly notices of the royal astronomical society}\ }\textbf
  {\bibinfo {volume} {473}},\ \bibinfo {pages} {271} (\bibinfo {year}
  {2018})},\ \Eprint {https://arxiv.org/abs/1709.03993} {arXiv:1709.03993
  [astro-ph.GA]} \BibitemShut {NoStop}%
\bibitem [{\citenamefont {{Joachimi}}\ \emph {et~al.}(2016)\citenamefont
  {{Joachimi}}, \citenamefont {{Gatuzz}}, \citenamefont {{Garc{\'\i}a}},\ and\
  \citenamefont {{Kallman}}}]{Joachimi_2016}%
  \BibitemOpen
  \bibfield  {author} {\bibinfo {author} {\bibfnamefont {K.}~\bibnamefont
  {{Joachimi}}}, \bibinfo {author} {\bibfnamefont {E.}~\bibnamefont
  {{Gatuzz}}}, \bibinfo {author} {\bibfnamefont {J.~A.}\ \bibnamefont
  {{Garc{\'\i}a}}},\ and\ \bibinfo {author} {\bibfnamefont {T.~R.}\
  \bibnamefont {{Kallman}}},\ }\bibfield  {title} {\bibinfo {title} {{On the
  detectability of CO molecules in the interstellar medium via X-ray
  spectroscopy}},\ }\href {https://doi.org/10.1093/mnras/stw1371} {\bibfield
  {journal} {\bibinfo  {journal} {Monthly notices of the royal astronomical
  society}\ }\textbf {\bibinfo {volume} {461}},\ \bibinfo {pages} {352}
  (\bibinfo {year} {2016})},\ \Eprint {https://arxiv.org/abs/1606.02285}
  {arXiv:1606.02285 [astro-ph.HE]} \BibitemShut {NoStop}%
\bibitem [{\citenamefont {{Kawamuro}}\ \emph {et~al.}(2020)\citenamefont
  {{Kawamuro}}, \citenamefont {{Izumi}}, \citenamefont {{Onishi}},
  \citenamefont {{Imanishi}}, \citenamefont {{Nguyen}},\ and\ \citenamefont
  {{Baba}}}]{Kawamuro_2020}%
  \BibitemOpen
  \bibfield  {author} {\bibinfo {author} {\bibfnamefont {T.}~\bibnamefont
  {{Kawamuro}}}, \bibinfo {author} {\bibfnamefont {T.}~\bibnamefont {{Izumi}}},
  \bibinfo {author} {\bibfnamefont {K.}~\bibnamefont {{Onishi}}}, \bibinfo
  {author} {\bibfnamefont {M.}~\bibnamefont {{Imanishi}}}, \bibinfo {author}
  {\bibfnamefont {D.~D.}\ \bibnamefont {{Nguyen}}},\ and\ \bibinfo {author}
  {\bibfnamefont {S.}~\bibnamefont {{Baba}}},\ }\bibfield  {title} {\bibinfo
  {title} {{AGN X-Ray Irradiation of CO Gas in NGC 2110 Revealed by Chandra and
  ALMA}},\ }\href {https://doi.org/10.3847/1538-4357/ab8b62} {\bibfield
  {journal} {\bibinfo  {journal} {The Astrophysical Journal}\ }\textbf
  {\bibinfo {volume} {895}},\ \bibinfo {eid} {135} (\bibinfo {year} {2020})},\
  \Eprint {https://arxiv.org/abs/2004.09394} {arXiv:2004.09394 [astro-ph.GA]}
  \BibitemShut {NoStop}%
\bibitem [{\citenamefont {Kincaid}\ and\ \citenamefont
  {Eisenberger}(1975)}]{Kincaid_1975}%
  \BibitemOpen
  \bibfield  {author} {\bibinfo {author} {\bibfnamefont {B.~M.}\ \bibnamefont
  {Kincaid}}\ and\ \bibinfo {author} {\bibfnamefont {P.}~\bibnamefont
  {Eisenberger}},\ }\bibfield  {title} {\bibinfo {title} {Synchrotron radiation
  studies of the $k$-edge photoabsorption spectra of kr, ${\mathrm{br}}_{2}$,
  and ge${\mathrm{cl}}_{4}$: A comparison of theory and experiment},\ }\href
  {https://doi.org/10.1103/PhysRevLett.34.1361} {\bibfield  {journal} {\bibinfo
   {journal} {Phys. Rev. Lett.}\ }\textbf {\bibinfo {volume} {34}},\ \bibinfo
  {pages} {1361} (\bibinfo {year} {1975})}\BibitemShut {NoStop}%
\bibitem [{\citenamefont {Iwayama}\ \emph {et~al.}(2020)\citenamefont
  {Iwayama}, \citenamefont {Nagasaka}, \citenamefont {Inoue}, \citenamefont
  {Owada}, \citenamefont {Yabashi},\ and\ \citenamefont
  {Harries}}]{Hiroshi_2020}%
  \BibitemOpen
  \bibfield  {author} {\bibinfo {author} {\bibfnamefont {H.}~\bibnamefont
  {Iwayama}}, \bibinfo {author} {\bibfnamefont {M.}~\bibnamefont {Nagasaka}},
  \bibinfo {author} {\bibfnamefont {I.}~\bibnamefont {Inoue}}, \bibinfo
  {author} {\bibfnamefont {S.}~\bibnamefont {Owada}}, \bibinfo {author}
  {\bibfnamefont {M.}~\bibnamefont {Yabashi}},\ and\ \bibinfo {author}
  {\bibfnamefont {J.~R.}\ \bibnamefont {Harries}},\ }\bibfield  {title}
  {\bibinfo {title} {Demonstration of transmission mode soft x-ray nexafs using
  third- and fifth-order harmonics of fel radiation at sacla bl1},\ }\bibfield
  {journal} {\bibinfo  {journal} {Applied Sciences}\ }\textbf {\bibinfo
  {volume} {10}},\ \href {https://doi.org/10.3390/app10217852}
  {10.3390/app10217852} (\bibinfo {year} {2020})\BibitemShut {NoStop}%
\bibitem [{\citenamefont {Hähner}(2006)}]{Hahner_2006}%
  \BibitemOpen
  \bibfield  {author} {\bibinfo {author} {\bibfnamefont {G.}~\bibnamefont
  {Hähner}},\ }\bibfield  {title} {\bibinfo {title} {Near edge x-ray
  absorption fine structure spectroscopy as a tool to probe electronic and
  structural properties of thin organic films and liquids},\ }\href
  {https://doi.org/10.1039/B509853J} {\bibfield  {journal} {\bibinfo  {journal}
  {Chem. Soc. Rev.}\ }\textbf {\bibinfo {volume} {35}},\ \bibinfo {pages}
  {1244} (\bibinfo {year} {2006})}\BibitemShut {NoStop}%
\bibitem [{\citenamefont {Hemraj-Benny}\ \emph {et~al.}(2006)\citenamefont
  {Hemraj-Benny}, \citenamefont {Banerjee}, \citenamefont {Sambasivan},
  \citenamefont {Balasubramanian}, \citenamefont {Fischer}, \citenamefont
  {Eres}, \citenamefont {Puretzky}, \citenamefont {Geohegan}, \citenamefont
  {Lowndes}, \citenamefont {Han}, \citenamefont {Misewich},\ and\ \citenamefont
  {Wong}}]{HemrajBenny_2006}%
  \BibitemOpen
  \bibfield  {author} {\bibinfo {author} {\bibfnamefont {T.}~\bibnamefont
  {Hemraj-Benny}}, \bibinfo {author} {\bibfnamefont {S.}~\bibnamefont
  {Banerjee}}, \bibinfo {author} {\bibfnamefont {S.}~\bibnamefont
  {Sambasivan}}, \bibinfo {author} {\bibfnamefont {M.}~\bibnamefont
  {Balasubramanian}}, \bibinfo {author} {\bibfnamefont {D.}~\bibnamefont
  {Fischer}}, \bibinfo {author} {\bibfnamefont {G.}~\bibnamefont {Eres}},
  \bibinfo {author} {\bibfnamefont {A.}~\bibnamefont {Puretzky}}, \bibinfo
  {author} {\bibfnamefont {D.}~\bibnamefont {Geohegan}}, \bibinfo {author}
  {\bibfnamefont {D.}~\bibnamefont {Lowndes}}, \bibinfo {author} {\bibfnamefont
  {W.}~\bibnamefont {Han}}, \bibinfo {author} {\bibfnamefont {J.}~\bibnamefont
  {Misewich}},\ and\ \bibinfo {author} {\bibfnamefont {S.}~\bibnamefont
  {Wong}},\ }\bibfield  {title} {\bibinfo {title} {Near-edge x-ray absorption
  fine structure spectroscopy as a tool for investigating nanomaterials},\
  }\href {https://doi.org/https://doi.org/10.1002/smll.200500256} {\bibfield
  {journal} {\bibinfo  {journal} {Small}\ }\textbf {\bibinfo {volume} {2}},\
  \bibinfo {pages} {26} (\bibinfo {year} {2006})},\ \Eprint
  {https://arxiv.org/abs/https://onlinelibrary.wiley.com/doi/pdf/10.1002/smll.200500256}
  {https://onlinelibrary.wiley.com/doi/pdf/10.1002/smll.200500256} \BibitemShut
  {NoStop}%
\bibitem [{\citenamefont {Dhez}\ \emph {et~al.}(2003)\citenamefont {Dhez},
  \citenamefont {Ade},\ and\ \citenamefont {Urquhart}}]{Dhez_2003}%
  \BibitemOpen
  \bibfield  {author} {\bibinfo {author} {\bibfnamefont {O.}~\bibnamefont
  {Dhez}}, \bibinfo {author} {\bibfnamefont {H.}~\bibnamefont {Ade}},\ and\
  \bibinfo {author} {\bibfnamefont {S.}~\bibnamefont {Urquhart}},\ }\bibfield
  {title} {\bibinfo {title} {Calibrated nexafs spectra of some common
  polymers},\ }\href
  {https://doi.org/https://doi.org/10.1016/S0368-2048(02)00237-2} {\bibfield
  {journal} {\bibinfo  {journal} {Journal of Electron Spectroscopy and Related
  Phenomena}\ }\textbf {\bibinfo {volume} {128}},\ \bibinfo {pages} {85}
  (\bibinfo {year} {2003})}\BibitemShut {NoStop}%
\bibitem [{\citenamefont {{Costantini}}\ \emph {et~al.}(2019)\citenamefont
  {{Costantini}}, \citenamefont {{Zeegers}}, \citenamefont {{Rogantini}},
  \citenamefont {{de Vries}}, \citenamefont {{Tielens}},\ and\ \citenamefont
  {{Waters}}}]{Costantini_2019}%
  \BibitemOpen
  \bibfield  {author} {\bibinfo {author} {\bibfnamefont {E.}~\bibnamefont
  {{Costantini}}}, \bibinfo {author} {\bibfnamefont {S.~T.}\ \bibnamefont
  {{Zeegers}}}, \bibinfo {author} {\bibfnamefont {D.}~\bibnamefont
  {{Rogantini}}}, \bibinfo {author} {\bibfnamefont {C.~P.}\ \bibnamefont {{de
  Vries}}}, \bibinfo {author} {\bibfnamefont {A.~G.~G.~M.}\ \bibnamefont
  {{Tielens}}},\ and\ \bibinfo {author} {\bibfnamefont {L.~B.~F.~M.}\
  \bibnamefont {{Waters}}},\ }\bibfield  {title} {\bibinfo {title} {{X-ray
  extinction from interstellar dust. Prospects of observing carbon, sulfur, and
  other trace elements}},\ }\href {https://doi.org/10.1051/0004-6361/201833820}
  {\bibfield  {journal} {\bibinfo  {journal} {Astronomy and Astrophysics}\
  }\textbf {\bibinfo {volume} {629}},\ \bibinfo {eid} {A78} (\bibinfo {year}
  {2019})},\ \Eprint {https://arxiv.org/abs/1906.08653} {arXiv:1906.08653
  [astro-ph.GA]} \BibitemShut {NoStop}%
\bibitem [{\citenamefont {Lee}\ \emph {et~al.}(2009)\citenamefont {Lee},
  \citenamefont {Xiang}, \citenamefont {Ravel}, \citenamefont {Kortright},\
  and\ \citenamefont {Flanagan}}]{Lee_2009}%
  \BibitemOpen
  \bibfield  {author} {\bibinfo {author} {\bibfnamefont {J.~C.}\ \bibnamefont
  {Lee}}, \bibinfo {author} {\bibfnamefont {J.}~\bibnamefont {Xiang}}, \bibinfo
  {author} {\bibfnamefont {B.}~\bibnamefont {Ravel}}, \bibinfo {author}
  {\bibfnamefont {J.}~\bibnamefont {Kortright}},\ and\ \bibinfo {author}
  {\bibfnamefont {K.}~\bibnamefont {Flanagan}},\ }\bibfield  {title} {\bibinfo
  {title} {{CONDENSED} {MATTER} {ASTROPHYSICS}: A {PRESCRIPTION} {FOR}
  {DETERMINING} {THE} {SPECIES}-{SPECIFIC} {COMPOSITION} {AND} {QUANTITY} {OF}
  {INTERSTELLAR} {DUST} {USING} x-{RAYS}},\ }\href
  {https://doi.org/10.1088/0004-637x/702/2/970} {\bibfield  {journal} {\bibinfo
   {journal} {The Astrophysical Journal}\ }\textbf {\bibinfo {volume} {702}},\
  \bibinfo {pages} {970} (\bibinfo {year} {2009})}\BibitemShut {NoStop}%
\bibitem [{\citenamefont {Wachulak}\ \emph {et~al.}(2018)\citenamefont
  {Wachulak}, \citenamefont {Duda}, \citenamefont {Bartnik}, \citenamefont
  {Sarzy\'{n}ski}, \citenamefont {Wegrzy\'{n}ski}, \citenamefont {Nowak},
  \citenamefont {Jancarek},\ and\ \citenamefont {Fiedorowicz}}]{Wachulak_2018}%
  \BibitemOpen
  \bibfield  {author} {\bibinfo {author} {\bibfnamefont {P.}~\bibnamefont
  {Wachulak}}, \bibinfo {author} {\bibfnamefont {M.}~\bibnamefont {Duda}},
  \bibinfo {author} {\bibfnamefont {A.}~\bibnamefont {Bartnik}}, \bibinfo
  {author} {\bibfnamefont {A.}~\bibnamefont {Sarzy\'{n}ski}}, \bibinfo {author}
  {\bibfnamefont {{\L}.}~\bibnamefont {Wegrzy\'{n}ski}}, \bibinfo {author}
  {\bibfnamefont {M.}~\bibnamefont {Nowak}}, \bibinfo {author} {\bibfnamefont
  {A.}~\bibnamefont {Jancarek}},\ and\ \bibinfo {author} {\bibfnamefont
  {H.}~\bibnamefont {Fiedorowicz}},\ }\bibfield  {title} {\bibinfo {title}
  {Compact system for near edge x-ray fine structure (nexafs) spectroscopy
  using a laser-plasma light source},\ }\href
  {https://doi.org/10.1364/OE.26.008260} {\bibfield  {journal} {\bibinfo
  {journal} {Opt. Express}\ }\textbf {\bibinfo {volume} {26}},\ \bibinfo
  {pages} {8260} (\bibinfo {year} {2018})}\BibitemShut {NoStop}%
\bibitem [{\citenamefont {Gissis}\ \emph {et~al.}(2020)\citenamefont {Gissis},
  \citenamefont {Behar}, \citenamefont {Fisher}, \citenamefont {Aricha},
  \citenamefont {Yeger}, \citenamefont {Avni},\ and\ \citenamefont
  {Schnitzer}}]{Gissis_Glider2020}%
  \BibitemOpen
  \bibfield  {author} {\bibinfo {author} {\bibfnamefont {I.}~\bibnamefont
  {Gissis}}, \bibinfo {author} {\bibfnamefont {E.}~\bibnamefont {Behar}},
  \bibinfo {author} {\bibfnamefont {A.}~\bibnamefont {Fisher}}, \bibinfo
  {author} {\bibfnamefont {S.}~\bibnamefont {Aricha}}, \bibinfo {author}
  {\bibfnamefont {E.}~\bibnamefont {Yeger}}, \bibinfo {author} {\bibfnamefont
  {U.}~\bibnamefont {Avni}},\ and\ \bibinfo {author} {\bibfnamefont
  {I.}~\bibnamefont {Schnitzer}},\ }\bibfield  {title} {\bibinfo {title}
  {Glider—a pulsed-current generator for laboratory astrophysics x-ray
  absorption experiments},\ }\href {https://doi.org/10.1063/1.5133056}
  {\bibfield  {journal} {\bibinfo  {journal} {Review of Scientific
  Instruments}\ }\textbf {\bibinfo {volume} {91}},\ \bibinfo {pages} {024701}
  (\bibinfo {year} {2020})},\ \Eprint
  {https://arxiv.org/abs/https://doi.org/10.1063/1.5133056}
  {https://doi.org/10.1063/1.5133056} \BibitemShut {NoStop}%
\bibitem [{\citenamefont {Berkowitz}(2015)}]{Berkowitz_2015}%
  \BibitemOpen
  \bibfield  {author} {\bibinfo {author} {\bibfnamefont {J.}~\bibnamefont
  {Berkowitz}},\ }\href {https://books.google.co.il/books?id=ae59BAAAQBAJ}
  {\emph {\bibinfo {title} {Atomic and Molecular Photoabsorption: Absolute
  Partial Cross Sections}}}\ (\bibinfo  {publisher} {Elsevier Science},\
  \bibinfo {year} {2015})\BibitemShut {NoStop}%
\bibitem [{\citenamefont {Arnaud}(1996)}]{Arnaud_1996}%
  \BibitemOpen
  \bibfield  {author} {\bibinfo {author} {\bibfnamefont {K.~A.}\ \bibnamefont
  {Arnaud}},\ }\bibfield  {title} {\bibinfo {title} {Astronomical data analysis
  software and systems v},\ }\href@noop {} {\bibfield  {journal} {\bibinfo
  {journal} {ASP Conf. Series}\ }\textbf {\bibinfo {volume} {101}},\ \bibinfo
  {pages} {11} (\bibinfo {year} {1996})}\BibitemShut {NoStop}%
\bibitem [{\citenamefont {Sham}\ \emph {et~al.}(1989)\citenamefont {Sham},
  \citenamefont {Yang}, \citenamefont {Kirz},\ and\ \citenamefont
  {Tse}}]{Sham_1989}%
  \BibitemOpen
  \bibfield  {author} {\bibinfo {author} {\bibfnamefont {T.~K.}\ \bibnamefont
  {Sham}}, \bibinfo {author} {\bibfnamefont {B.~X.}\ \bibnamefont {Yang}},
  \bibinfo {author} {\bibfnamefont {J.}~\bibnamefont {Kirz}},\ and\ \bibinfo
  {author} {\bibfnamefont {J.~S.}\ \bibnamefont {Tse}},\ }\bibfield  {title}
  {\bibinfo {title} {K-edge near-edge x-ray-absorption fine structure of
  oxygen- and carbon-containing molecules in the gas phase},\ }\href
  {https://doi.org/10.1103/PhysRevA.40.652} {\bibfield  {journal} {\bibinfo
  {journal} {Phys. Rev. A}\ }\textbf {\bibinfo {volume} {40}},\ \bibinfo
  {pages} {652} (\bibinfo {year} {1989})}\BibitemShut {NoStop}%
\bibitem [{\citenamefont {Hasoglu}\ \emph {et~al.}(2010)\citenamefont
  {Hasoglu}, \citenamefont {Abdel-Naby}, \citenamefont {Gorczyca},
  \citenamefont {Drake},\ and\ \citenamefont {McLaughlin}}]{Hasoglu_2010}%
  \BibitemOpen
  \bibfield  {author} {\bibinfo {author} {\bibfnamefont {M.~F.}\ \bibnamefont
  {Hasoglu}}, \bibinfo {author} {\bibfnamefont {S.~A.}\ \bibnamefont
  {Abdel-Naby}}, \bibinfo {author} {\bibfnamefont {T.~W.}\ \bibnamefont
  {Gorczyca}}, \bibinfo {author} {\bibfnamefont {J.~J.}\ \bibnamefont
  {Drake}},\ and\ \bibinfo {author} {\bibfnamefont {B.~M.}\ \bibnamefont
  {McLaughlin}},\ }\bibfield  {title} {\bibinfo {title} {K-{SHELL}
  {PHOTOABSORPTION} {STUDIES} {OF} {THE} {CARBON} {ISONUCLEAR} {SEQUENCE}},\
  }\href {https://doi.org/10.1088/0004-637x/724/2/1296} {\bibfield  {journal}
  {\bibinfo  {journal} {The Astrophysical Journal}\ }\textbf {\bibinfo {volume}
  {724}},\ \bibinfo {pages} {1296} (\bibinfo {year} {2010})}\BibitemShut
  {NoStop}%
\bibitem [{\citenamefont {{Garc{\'\i}a}}\ \emph {et~al.}(2009)\citenamefont
  {{Garc{\'\i}a}}, \citenamefont {{Kallman}}, \citenamefont {{Witthoeft}},
  \citenamefont {{Behar}}, \citenamefont {{Mendoza}}, \citenamefont
  {{Palmeri}}, \citenamefont {{Quinet}}, \citenamefont {{Bautista}},\ and\
  \citenamefont {{Klapisch}}}]{Garcia_2009}%
  \BibitemOpen
  \bibfield  {author} {\bibinfo {author} {\bibfnamefont {J.}~\bibnamefont
  {{Garc{\'\i}a}}}, \bibinfo {author} {\bibfnamefont {T.~R.}\ \bibnamefont
  {{Kallman}}}, \bibinfo {author} {\bibfnamefont {M.}~\bibnamefont
  {{Witthoeft}}}, \bibinfo {author} {\bibfnamefont {E.}~\bibnamefont
  {{Behar}}}, \bibinfo {author} {\bibfnamefont {C.}~\bibnamefont {{Mendoza}}},
  \bibinfo {author} {\bibfnamefont {P.}~\bibnamefont {{Palmeri}}}, \bibinfo
  {author} {\bibfnamefont {P.}~\bibnamefont {{Quinet}}}, \bibinfo {author}
  {\bibfnamefont {M.~A.}\ \bibnamefont {{Bautista}}},\ and\ \bibinfo {author}
  {\bibfnamefont {M.}~\bibnamefont {{Klapisch}}},\ }\bibfield  {title}
  {\bibinfo {title} {Nitrogen k-shell photoabsorption},\ }\href
  {https://doi.org/10.1088/0067-0049/185/2/477} {\bibfield  {journal} {\bibinfo
   {journal} {The Astrophysical Journal Supplement Series}\ }\textbf {\bibinfo
  {volume} {185}},\ \bibinfo {pages} {477} (\bibinfo {year}
  {2009})}\BibitemShut {NoStop}%
\bibitem [{\citenamefont {Hitchcock}\ and\ \citenamefont
  {Brion}(1980)}]{Hitchcock_1980}%
  \BibitemOpen
  \bibfield  {author} {\bibinfo {author} {\bibfnamefont {A.}~\bibnamefont
  {Hitchcock}}\ and\ \bibinfo {author} {\bibfnamefont {C.}~\bibnamefont
  {Brion}},\ }\bibfield  {title} {\bibinfo {title} {K-shell excitation spectra
  of co, n2 and o2},\ }\href
  {https://doi.org/https://doi.org/10.1016/0368-2048(80)80001-6} {\bibfield
  {journal} {\bibinfo  {journal} {Journal of Electron Spectroscopy and Related
  Phenomena}\ }\textbf {\bibinfo {volume} {18}},\ \bibinfo {pages} {1 }
  (\bibinfo {year} {1980})}\BibitemShut {NoStop}%
\bibitem [{\citenamefont {McLaughlin}\ \emph {et~al.}(2013)\citenamefont
  {McLaughlin}, \citenamefont {Ballance}, \citenamefont {Bowen}, \citenamefont
  {Gardenghi},\ and\ \citenamefont {Stolte}}]{McLaughlin_2013}%
  \BibitemOpen
  \bibfield  {author} {\bibinfo {author} {\bibfnamefont {B.~M.}\ \bibnamefont
  {McLaughlin}}, \bibinfo {author} {\bibfnamefont {C.~P.}\ \bibnamefont
  {Ballance}}, \bibinfo {author} {\bibfnamefont {K.~P.}\ \bibnamefont {Bowen}},
  \bibinfo {author} {\bibfnamefont {D.~J.}\ \bibnamefont {Gardenghi}},\ and\
  \bibinfo {author} {\bibfnamefont {W.~C.}\ \bibnamefont {Stolte}},\ }\bibfield
   {title} {\bibinfo {title} {{HIGH} {PRECISION} k-{SHELL} {PHOTOABSORPTION}
  {CROSS} {SECTIONS} {FOR} {ATOMIC} {OXYGEN}: {EXPERIMENT} {AND} {THEORY}},\
  }\href {https://doi.org/10.1088/2041-8205/771/1/l8} {\bibfield  {journal}
  {\bibinfo  {journal} {The Astrophysical Journal}\ }\textbf {\bibinfo {volume}
  {771}},\ \bibinfo {pages} {L8} (\bibinfo {year} {2013})}\BibitemShut
  {NoStop}%
\bibitem [{\citenamefont {Huebner}\ and\ \citenamefont
  {Mukherjee}(2015)}]{Huebner_2015}%
  \BibitemOpen
  \bibfield  {author} {\bibinfo {author} {\bibfnamefont {W.}~\bibnamefont
  {Huebner}}\ and\ \bibinfo {author} {\bibfnamefont {J.}~\bibnamefont
  {Mukherjee}},\ }\bibfield  {title} {\bibinfo {title} {Photoionization and
  photodissociation rates in solar and blackbody radiation fields},\ }\href
  {https://doi.org/https://doi.org/10.1016/j.pss.2014.11.022} {\bibfield
  {journal} {\bibinfo  {journal} {Planetary and Space Science}\ }\textbf
  {\bibinfo {volume} {106}},\ \bibinfo {pages} {11} (\bibinfo {year}
  {2015})}\BibitemShut {NoStop}%
\end{thebibliography}%
		
\end{document}